\documentstyle[12pt]{article}
\def\al{\alpha}
\def\be{\beta}
\def\ga{\gamma}

\def\ep{\epsilon}

\def\ka{\kappa}
\def\la{\lambda}

\def\si{\sigma}

\def\ps{\psi}

\def\Ga{\Gamma}

\def\mn{{\mu\nu}}

\def\cl{{\cal L}}

\def\half{{\textstyle{1\over 2}}}
\def\frac#1#2{{\textstyle{{#1}\over {#2}}}}

\def\lsim{\mathrel{\rlap{\lower4pt\hbox{\hskip1pt$\sim$}}
    \raise1pt\hbox{$<$}}}
\def\gsim{\mathrel{\rlap{\lower4pt\hbox{\hskip1pt$\sim$}}
    \raise1pt\hbox{$>$}}}
\def\sqr#1#2{{\vcenter{\vbox{\hrule height.#2pt
         \hbox{\vrule width.#2pt height#1pt \kern#1pt
         \vrule width.#2pt}
         \hrule height.#2pt}}}}

\def\lrprt{\stackrel{\leftrightarrow}{\partial}}
\def\lrprtnu{\stackrel{\leftrightarrow}{\partial^\nu}}

\def\lrprtmuupper{\stackrel{\leftrightarrow}{\partial^\mu}}

\def\lrDnu{\stackrel{\leftrightarrow}{D^\nu}}

\newcommand{\beq}{\begin{equation}}
\newcommand{\eeq}{\end{equation}}
\newcommand{\bea}{\begin{eqnarray}}
\newcommand{\eea}{\end{eqnarray}}
\newcommand{\rf}[1]{(\ref{#1})}
 
\renewenvironment{thebibliography}[1]
 { \rm
   \begin{list}{\arabic{enumi}.}
    {\usecounter{enumi} \setlength{\parsep}{0pt}
     \setlength{\itemsep}{3pt} \settowidth{\labelwidth}{#1.}
     \sloppy
    }}{\end{list}}
 
\begin{document}
\titlepage
 

            
\begin{center}
{{\bf REDEFINING SPINORS IN LORENTZ-VIOLATING QED \\}
\vglue 1.0cm
{Don Colladay and Patrick McDonald\\} 
\bigskip
{\it New College of Florida\\}
\medskip
{\it Sarasota, FL, 34243, U.S.A.\\}
 
\vglue 0.8cm
}
\vglue 0.3cm
 
\end{center}
 
{\rightskip=3pc\leftskip=3pc\noindent
An analysis of spinor redefinitions in the context of the
Lorentz-violating QED extension is performed.  Certain parameters
that apparently violate Lorentz invariance are found to be physically
irrelevant as they can be removed from the lagrangian using an
appropriate redefinition of the spinor field components.  It is shown
that conserved currents may be defined using a modified action of the
complex extension of the Lorentz group on the redefined spinors.  This
implies a natural correspondence between the apparently
Lorentz-violating theory and conventional QED.  
Redefinitions involving derivatives are shown to relate certain
terms in the QED extension to lagrangians involving nonlocal
interactions or skewed coordinate systems.  The redundant parameters in
the QED extension are identified and the lagrangian is rewritten in
terms of physically relevant coupling constants.  The resulting
lagrangian contains only physically relevant parameters and transforms
conventionally under Lorentz transformations.}

\vskip 1 cm

\vskip 1 cm


\newpage
 
\baselineskip=20pt
 
{\bf \noindent I. INTRODUCTION}
\vglue 0.4cm

The possibility of miniscule violations of Lorentz invariance arising
from a more fundamental theory of nature has been of recent interest
\cite{cpt98}.
For example, such violations may arise in the low-energy limit of
string theory
\cite{kps}, or physically realistic noncommutative field theories
\cite{noncom}.   
The full standard model extension uses 
the general concept of spontaneous symmetry breaking to
construct a lagrangian consisting of all possible terms
involving standard-model fields
that are observer Lorentz scalars,
including terms having coupling coefficients with Lorentz indices.
At low energies,
the relevant operators that are gauge invariant all have mass dimension
$D\le 4$, and are given in \ \cite{ck}.
At higher energy scales nonrenormalizable terms are expected to play a
role in the theoretical consistency of the model \cite{kle}.

Various experiments have placed stringent bounds on
parameters in the standard-model extension,
including
comparative tests of quantum electrodynamics (QED)
in Penning traps and colliders \cite{bkr,gg,hd,rm,scatt,pick},
spectroscopy of hydrogen and antihydrogen \cite{bkr2,dp},
measurements of muon properties
\cite{bkl,vh},
clock-comparison experiments
\cite{ccexpt,kla,lh,db},
observations of the behavior of a spin-polarized torsion pendulum
\cite{bk,bh},
measurements of cosmological birefringence \cite{cfj,ck,jk,pvc,mewes},
studies of neutral-meson oscillations
\cite{kpcvk,k99,bexpt},
and observations of the baryon asymmetry \cite{bckp}.

In the theoretical results involving experimentally observable
quantities, some of the parameters in the standard model extension do
not appear while others occur only in specific linear combinations.
The reason behind this is that some parameters that apparently violate
Lorentz invariance when the spinor field is assumed to transform in
the standard way under the action of the Lorentz group do not in fact violate
this symmetry when the action on the field is appropriately modified such
that the associated Lorentz currents are conserved. 
The freedom to select spinor coordinates in different ways generates a
natural equivalence relation on the collection of lagrangians. 
Different lagrangians in the same equivalence class are related by
field redefinitions; that is, by an invertible map between fields used
to describe the same physics.   The explicit construction of the
redefinition used in this paper appears in Eq.\rf{fredef}. All
lagrangians in the same class are physically equivalent and 
the spinor transformation properties can be implemented so that
the Lorentz currents are as close to conserved as possible.  
In other words, redefinitions may be used to define the currents 
so that they absorb the apparently Lorentz-violating terms which are 
obstructions to conservation.
This means that one can use the
transformation properties to eliminate a subset of the parameters 
appearing in the standard model extension and no more.

In this paper, the effects of field redefinitions in the context of
extended QED are examined in detail.  Particular terms in
the standard model extension are already known to be unobservable since explicit
redefinitions of the spinor components have previously been considered
\cite{ck}.  It is the goal of this work to analyze a more general
set of field redefinitions and use them to simplify the full
Lorentz-violating lagrangian as much as possible.  The basic idea is
to remove parameters that depend explicitly on the spinor coordinates.
Once these redundant parameters have been eliminated, the remaining
field transforms according to the standard action of the complex
Lorentz group.  

The paper is organized as follows.  In section II, the extended QED
theory is summarized.  In Section III, an analysis is presented of the
field redefinitions that are used to generate specific terms in the QED
extension.  The effects of transformations which do not include terms
involving differentiation as well as those that do are investigated.
The currents associated with U(1) and Poincar\'e group transformations
for the general QED extension are derived in section IV.  
It is shown that conserved currents can be defined when only redundant
parameters are present by using a similar representation to the
conventional complex Lorentz group action.  Section V
contains the construction of the physical extended QED lagrangian with
all redundant parameters removed.

\vglue 0.6cm
{\bf \noindent II. EXTENDED QED}
\vglue 0.4cm

To study the effects of specific field redefinitions, we restrict our
attention to the QED subset involving only the electron and photon
sectors of the full standard model extension presented in reference
\cite{ck}.   In the pure-photon sector,
there is one CPT-even ($k_F$) and one CPT-odd
($k_{AF}$) Lorentz-violating term.  The free photon lagrangian is 
\beq
\cl_{\rm photon} = - \frac 1 4 F_{\mu\nu}F^{\mu\nu}
-\frac 1 4 (k_F)_{\ka\la\mu\nu} F^{\ka\la}F^{\mu\nu}
+ \half (k_{AF})^\ka \ep_{\ka\la\mu\nu} A^\la F^{\mu\nu} 
\quad ,
\label{lorqed}
\eeq
where the coupling $k_F$ is a real, dimensionless coupling that
can be taken to have the symmetries of the Riemann tensor, and the
coefficient $k_{AF}$
is real and has dimensions of mass.

Denoting the four-component electron field by $\ps$
and the electron mass by $m$,
the general QED lagrangian for electrons and photons 
including Lorentz-violating interactions arising from a 
generic spontaneous symmetry breaking mechanism is
\beq
\label{lorviol}
\cl^{\rm QED}_{\rm electron} = 
\frac i 2 \overline{\ps} \Gamma_\nu \lrDnu \ps 
- \overline{\ps} M \ps
\label{genlag}
\quad ,
\eeq
where $D_\mu = \partial_\mu + i q A_\mu$ is the usual covariant
derivative, 
\beq
\Gamma_{\nu} = \ga_{\nu} + c_{\mn}\ga^{\mu} + 
d_{\mu\nu} \ga_5 \ga^\mu +e_\nu +if_\nu\gamma_5+\half g_{\lambda
\mu \nu}\sigma^{\lambda \mu}
\quad ,
\label{gammanu}
\eeq
and
\beq
M = m + i m_5 \gamma_5+a_{\mu } \ga^{\mu} + b_{\mu} \ga_5 \ga^{\mu}
+ \half H_{\mu\nu}\si^{\mu\nu}
\quad .
\eeq
Note that any Lorentz-{\it preserving} terms that arise
from spontaneous symmetry breaking can be absorbed into the
bare mass terms 
$m$, $m_5$, and the overall normalization of the lagrangian.  The
normalization is chosen such that 
the coefficient of the $\gamma_\nu$ term in Eq.\rf{gammanu} is one.
The coupling coefficients
$a$, $b$, $c$, $d$, $e$, $f$, $g$, $m_5$, and $H$
are real, constant parameters related to the vacuum expectation value 
of contributing tensor fields in the underlying theory.

Some of these parameters (or combinations of parameters) are only
apparently Lorentz-violating as the lagrangian containing them can be
shown to be equivalent to the standard lagrangian using the appropriate
field redefinition.  This question is taken up in the next section.

\vglue 0.6cm
{\bf \noindent III. GENERAL FERMION FIELD REDEFINITIONS}
\vglue 0.4cm

Some of the Lorentz-violating couplings in Eq.\rf{lorviol} can in
fact be eliminated through a redefinition of the spinor field.  
To determine precisely which terms can be removed in this manner, it
is useful to begin
with the standard Dirac lagrangian in terms of $\psi$ with no
Lorentz-violating terms and perform a field redefinition of the form
$\psi = R
\chi$, where $R$ is some operator.  The lagrangian in terms of $\chi$
will contain terms included in the full Lorentz-violating lagrangian.  
In this section, we examine possible choices for the field
redefinition and examine the resulting terms in the lagrangian.

To see which terms can be removed from the
theory, we consider generic redefinitions of the
fermion fields of the form
\beq
\label{fredef}
\psi(x) = [1 + f(x,\partial)] \chi(x) \quad ,
\eeq
where $f(x,\partial)$ represents a general $4 \times 4$ matrix function
of the coordinates and derivatives.
Only lowest order terms in the field redefinition parameters
are retained
since the Lorentz-violating couplings in the full lagrangian
are assumed small.  
By applying this transformation to the conventional free fermion
lagrangian (containing no Lorentz-violating parameters) we will see
which terms can be eliminated from the extended theory by applying the
inverse transformation.

To simplify the task, note that the Lorentz-violating terms generated
by this transformation must have no explicit dependence
on the coordinates and must consist of dimension
$D \le 4$ operators\footnote{Actually, terms in the
transformed lagrangian may contain an explicit dependence on $x$ or may be
of dimension greater than four provided that the terms are total
divergences and therefore can be removed from the action. For examples,
see the relevant terms in Eq.\rf{chilag}.}.   Candidate terms for 
$f(x,\partial)$ up to second order in $x$ and $\partial$ are of the
form
\beq
f(x,\partial) = v \cdot \Gamma +
i\theta + iA_\mu x^\mu + B_\mu \partial^\mu +
\ga_5 \tilde B_\mu \partial ^\mu + C_{\mu\nu}x^\mu \partial^\nu
\quad ,
\label{translag}
\eeq
where $v$ represents a set of arbitrary complex constants multiplying
an arbitrary gamma matrix, denoted $\Gamma,$  in the set $m\{i \gamma_5,
\gamma^\mu, \gamma_5 \gamma^\mu, \sigma^{\mu\nu} \}$, $\theta$ is a
complex constant, while $A_\mu$, $B_\mu$, $\tilde B_\mu$, and
$C_{\mu\nu}$ are arbitrary real constants.  
Note that this is a generalization of the field redefinitions
previously considered in \cite{ck}.

The terms
Re$\theta$,
$B_\mu$ and the antisymmetric part of
$C_{\mu
\nu}$ (together with the appropriate spin components $v_{\mn}$) are
simply the generators of the $U(1)$ and Poincar\'e groups and
are symmetries of the conventional lagrangian.  These terms do not
generate any artificial Lorentz-violating parameters.  The term
Im$\theta$ rescales the lagrangian and can be absorbed into the
other constants.  This leaves several independent transformations that may
generate artificial Lorentz-violating terms.  We proceed to calculate
these explicitly in the rest of this section.  Since we are working to
lowest order in Lorentz-violating parameters, we can consider each
term independently.

First, we summarize the results obtained using the $v$ terms which
have been previously described in \cite{ck}.   An explicit example is
presented to illustrate the general method. As is shown in the next
section, a field redefinition of this type can be interpreted as selecting a
new basis in spinor space for the representations of SL(2,C), the
complex extension of the Lorentz group. The standard lagrangian
expressed in terms of the redefined field is given by
\bea
\label{vdotgamlag}
\cl & = & \frac i 2 \overline \psi \gamma^\mu \lrprt_\mu
\psi  -m \overline \psi \psi \nonumber \\
& = & \cl_0+\frac i 2  \overline \chi [\{\gamma^\mu,\Gamma\cdot {\rm
Re} v\} + i [\gamma^\mu,\Gamma \cdot {\rm Im} v]]\lrprt_\mu \chi 
- 2 m {\rm Re} v \cdot \overline \chi \Gamma \chi
\quad ,
\eea
where $\cl_0$ is the conventional free field lagrangian in terms of
$\chi$. For example, consider the field redefinition induced by
$v \cdot \Gamma = v_\mu
\gamma^\mu$.  Using the above relation yields
\beq
\label{aeredef}
\cl = \cl_0 + {\rm Re} v_\mu [i \overline \chi \lrprtmuupper \chi - 2 m
\overline \chi \gamma^\mu \chi]- i {\rm Im} v_\mu [\overline \chi
\sigma^{\mu \nu} \lrprt_\nu \chi].
\eeq
Inspection of the term proportional to ${\rm Im}v_\mu$ indicates that
the four terms in the extended lagrangian \rf{genlag} of the form 
\beq
g_{\lambda \mu \nu} = 2{\rm Im} (v_\mu g_{\la \nu} - v_\la
g_{\mu \nu})
\quad ,
\eeq
do not contribute in lowest order to the free fermion lagrangian.
Examination of the terms multiplying ${\rm Re} v_\mu$ indicates that
the simultaneous choice of
\beq
e^\mu = 2 {\rm Re} v^\mu \quad , \quad a^\mu = 2 m {\rm Re} v^\mu
\eeq    
can be removed from the lagrangian.  This means that the field
redefinition can remove either $e^\mu$ or $a^\mu$, but not both,
unless $a^\mu = m e^\mu$ happens to hold in the original
lagrangian.
Similar calculations can be done for the other choices of $v
\cdot \Gamma$.  The results are summarized in table 1.
Note that a (finite) transformation of the form $e^{i v \gamma_5}$ with
$v \in \Re$ is used to remove any term of the form $m_5$ in the
original lagrangian to all orders.  The effect is an $m_5$ dependent
mixing of some of the Lorentz-violating parameters, but the structure
is essentially unchanged.

\begin{table}{}
\begin{tabular}{||c|l|l||}
\hline \hline
$\bf v \cdot \Gamma$ & $\bf v \in \Re $\quad ($v \equiv $ Re {\bf v}) &
$\bf v
\in
\Im$\quad  ($v \equiv $ Im {\bf{v}}) \\
\hline
$v(i \gamma_5)$ & Used to eliminate $m_5$ term & 
$d_{\mu\nu} = 2 v g_{\mu\nu}$ \\ \hline
$v_\mu \gamma^\mu$ & $e_\mu = 2v_\mu  \quad {\rm and} \quad a_\mu = 2m
v_\mu$ & $g_{\lambda \mu \nu} = 2( v_\mu g_{\lambda \nu} -
v_\lambda g_{\mn})
$ \\ \hline
$v_\mu \gamma_5 \gamma^\mu$ & $g_{\lambda \mu \nu} =
-2\epsilon_{\lambda \mu \nu}^{~~~~\alpha} v_\alpha \quad {\rm and}
\quad b_\mu = 2 m v_\mu$ & $f_\mu = -2 v_\mu$ \\ \hline
$v_{\mn} \sigma^\mn$ & $d_{\mn} = - 2 \epsilon_\mn^{~~~\alpha \beta}
v_{\alpha \beta} \quad {\rm and} \quad \frac{1} {2} H_\mn = 2 m v_\mn$
&
$c_{\mn} = 2 v_{[\mn]}$ \\ \hline \hline
\end{tabular}
\caption{Summary of terms generated by field redefinitions of the form
$v \cdot \Gamma$.  }
\end{table}

Next, we consider the $A$, $B$, $\tilde B$, and $C$ redefinitions.  To
lowest order in these parameters, the transformed lagrangian becomes
\arraycolsep=1pt
\bea
\cl = & & \cl_0 - \overline \chi A_\mu \gamma^\mu \chi + B_\mu
\partial^\mu \cl_0 - \tilde B_\mu \partial^\mu[\frac i 2 \overline \chi
\ga_5 \gamma^\nu \lrprt_\nu\chi ] -  m \overline \chi \ga_5 \tilde
B_\mu
\lrprtmuupper \chi \nonumber \\
&  & +C_\mn x^\mu
\partial^\nu
\cl_0 + \frac i 2 C_\mn \overline \chi \gamma^\mu \lrprtnu \chi
\quad .
\label{chilag}
\eea 
The $A$ term can be used to eliminate $a^\mu$ as is discussed in
detail in reference \cite{ck}.  The
$B$ term is a total divergence that drops out of the action.  This is a
direct consequence of translational invariance since under a (finite)
translation of the coordinates by $B$
\beq
\psi(x) = e^{B \cdot \partial} \chi (x) = \chi(x + B) = \chi(x^\prime)
\quad ,
\eeq
and the action takes the same form in the translated coordinate
system. The first $C$ term can
be partially integrated to yield a total divergence and a rescaling of
$\cl_0$.  The final $C$ term is of the form $c$ as defined in
Eq.\rf{genlag}.  

A few remarks are in order regarding the above transformation
involving $C$. First, note that such a
field redefinition appears equivalent to changing fermion coordinates
to a system with a new (constant) metric that skews the coordinates. 
We can see this by examination of the transformation
\beq
\label{credef}
\psi(x) = (1 + C^\mn x_\mu \partial_\nu)\chi(x)\approx e^{C^\mn x_\mu
\partial_\nu} \chi(x) =
\chi(x + C \cdot x) = \chi(x^\prime) \quad ,
\eeq
where $x^{\prime \mu} = x^\mu + C^\mu_{~\nu} x^\nu$ are the new field 
coordinates.  
This redefinition is therefore equivalent to transforming to a skewed
coordinate system with a nondiagonal metric given by
\beq
g^{\prime \mn} = \eta^\mn + C^{(\mn)} 
\quad .
\eeq 
Rewriting the transformed lagrangian in terms of this new metric
yields 
\beq
\cl = \frac i 2 \overline \chi(x^\prime) \tilde \gamma^\mu
{\stackrel{\leftrightarrow}{\partial^{\,\prime}_\mu}}
\chi(x^\prime) - m \overline \chi(x^\prime)
\chi(x^\prime)
\eeq
where the modified matrices $\tilde \gamma^\mu = (\eta^\mu_{~\nu} +
c^\mu_{~\nu})
\gamma^\nu$ satisfy the relations $\{ \tilde \gamma^\mu,\tilde
\gamma^\nu\} = 2g^{\prime \mn}$.
The resulting lagrangian can be related to the conventional one using
the vierbein formalism of general relativity by performing the
appropriate general coordinate transformation. This shows that there
is a natural association between the theory containing a $c$ term and
a theory formulated in a skewed coordinate system defined by the
metric given above.  If the free fermions are the only component to
the theory it is possible to perform the appropriate general
coordinate transformation on the skewed coordinates relating it to the
conventional case.  This is because it is not possible to distinguish
the theory in a skewed coordinate system (with a $c$ term) from a
conventional theory in an orthonormal system since fermion
propagation properties are the only tool available to define the
coordinate system itself.

However, when photon interactions are incorporated through
the covariant derivative, it is no longer possible to perform the
general coordinate transformation without affecting the photon sector.
There is now an alternate way to distinguish the coordinates physically
so that the skewed coordinates become observable. Under the fermion
field redefinition, the fermion-photon interaction term becomes
\beq
\cl_{int} = 
- q \overline \psi(x) \gamma^\mu A_\mu(x)\psi(x) 
\longrightarrow 
- q \overline \chi (x^\prime) \tilde \gamma^\mu A^{\prime}_{\mu} (x)
\chi(x^\prime)
\eeq
where the photon field is expressed as a function of the conventional
coordinates $x$, but its components are resolved in the
modified coordinates $x^\prime$. 
In this picture, the theory exhibits a form of nonlocality since the fermion fields
interact with the photon field at different spacetime points.
If the
photon field is re-expressed in terms of the new coordinates
$x^\prime$, 
the lagrangian becomes local, but picks up an extra term that breaks
the natural association between the theories.  
Yet another approach is to redefine the physical photon field
$A_\mu^\prime(x)
\rightarrow A_\mu^\prime(x^\prime)$,
but the new metric introduces corrections of the form in Eq.\rf{lorqed}
into the kinetic photon sector.  Therefore, the
photon interactions prevent the trivial elimination of symmetric $c$
terms using the above field redefinition.  Similar problems arise when
using other derivative transformations, so these are not
considered in detail in subsequent sections of this paper. For example,
the term
$\tilde B$ in Eq.\rf{translag} generates a term of the form $f$
(defined in Eq.\rf{genlag}) as well as a total divergence in the
transformed lagrangian of Eq.\rf{chilag}.  This transformation
corresponds to a shift in opposite directions for the left-handed and
right-handed fields due to the presence of
$\gamma_5$.  There is therefore a natural correspondence between the
lagrangian with an $f$ term and the conventional theory provided the
left-handed and right-handed fields can be translated
independently\footnote{Actually, one of the redefinitions listed in
table 1 can also be used to eliminate the $f$ term, so it is always
removable from the lagrangian.}.  The interaction that breaks the
correspondence with the conventional theory in this case is the
fermion mass term that mixes left-handed and right-handed fields. A
similar situation occurs when the $C$ transformation discussed above
is multiplied by
$\gamma_5$.  Symmetric components of the form
$d$ defined in Eq.\rf{genlag} arise in the transformed lagrangian, but
the mass term depends explicitly on $x$, therefore
breaking the natural correspondence between the theories.

Finite field transformations can be constructed through
exponentiation of Eq.\rf{fredef}.  The results of these
transformations are often much more complicated than the infinitesimal
ones since several parameters can be mixed.  As
an example, consider a field redefinition of the form
\beq
\psi = e^{v_\mu \gamma^\mu} \chi = (\cosh{v} + {v_\mu
\gamma^\mu\over v} \sinh{v})\chi \quad ,
\label{finitefredef}
\eeq
with $v_\mu$ real, timelike, and the quantity $v$ defined by $v =
\sqrt{v_\mu v^\mu}$. Application of this transformation to the
standard lagrangian in terms of $\psi$ yields a lagrangian for $\chi$
with apparent Lorentz-violating parameters and a modified mass given by
\bea
e^\mu = {v^\mu \over v} \sinh{2v} \quad &;& \quad c^\mn = {v^\mu v^\nu
\over v^2}(\cosh{2v} - 1)\quad , \nonumber \\
m^\prime = m \cosh{2v} \quad &;& \quad a^\mu = {m v^\mu \over
v}\sinh{2v} \quad .
\eea
Note that the corrections to $c$ and $m$ terms appear only at second
order in
$v$.  

Such a choice of parameters in the QED
extension leaves the dispersion relation unaltered and therefore
leads to no stability problems or microcausality violations.  This is
true for any finite field redefinition of the form $e^{v \cdot
\Gamma}$ since the field redefinition commutes with the square of the
conventional Dirac equation.  
The set of all such transformations generates a class of lagrangians
equivalent to the conventional one.

The derivative field redefinitions may also be exponentiated to yield
finite transformations.  In the cases of $\tilde B$ and $C$ of
Eq.\rf{translag}, finite transformations correspond to finite
coordinate transformations, possibly different for various spinor
components.  
As in the infinitessimal case, interactions between various spinor
components and other fields limit the usefulness of these
transformations due to nonlocality problems.

To gain further insight into the
invariance of the physics under the  above transformations, it is
useful to compute the currents associated with the generators of the
Poincar\'e group. We will see that it is necessary to redefine the action of
the complex Lorentz group along with the field 
in order to yield maximally
conserved Lorentz generators.

\vglue 0.6cm
{\bf \noindent IV. POINCAR\'E GENERATORS}
\vglue 0.4cm

We start with a general free fermion lagrangian of the form
\beq
\cl = \frac i 2 \overline \chi \Gamma^\nu \lrprt_\nu \chi -
\overline \chi M \chi
\quad ,
\eeq
and apply Noether's theorem to obtain the divergence of the currents
associated with various continuous transformations of the field.  
Invariance under a global $U(1)$ phase transformation yields a
conserved current of
\beq
\label{4cur}
j^\mu = \overline \chi \Gamma^\mu \chi
\quad ,
\eeq
satisfying $\partial_\mu j^\mu =0$.
Invariance under translations yields a conserved energy momentum
tensor of
\beq
\label{emom}
\Theta^\mn = \frac i 2 \overline \chi \Gamma^\mu \lrprtnu \chi
\quad ,
\eeq
satisfying $\partial_\mu \Theta^\mn = 0$.  

The lagrangian is no longer invariant under the conventional
action of the Lorentz group, so the divergences of the corresponding
currents will not vanish.  These can be calculated using the standard
technique of writing the action over an arbitrary four-volume in terms
of boosted coordinates and fields at
$x^{\prime\mu} =
\Lambda^\mu_{~\nu} x^\nu
\approx x^\mu +
\epsilon^\mu_{~\nu} x^\nu$ and calculating the variation to lowest
order in $\ep^\mu_{~\nu}$. 
A choice must be made for the induced transformation properties of the
spinor components of $\chi$ under the complex extension of the Lorentz
group $SL(2,C)$. Using the standard $S(\Lambda) = 1 - \frac i 4
\si_\mn \ep^\mn$ yields currents given by
\beq
\partial_\alpha j^{\alpha \mn} = X^{\mn}
\quad ,
\eeq
where 
\beq
j^{\alpha \mn} = x^{[\mu} \Theta^{\alpha \nu]} 
+ \frac 1 4 \overline \chi \{ \Gamma^\alpha , \sigma^\mn\}\chi
\quad ,
\label{lorcur}
\eeq
and 
\bea
\nonumber
X^\mn & = & - a^{[\mu} \overline \chi \gamma^{\nu]} \chi 
- b^{[\mu} \overline \chi \gamma_5 \gamma^{\nu]} \chi 
-\frac i 2 \overline \chi (c^{[\nu \alpha}\gamma^{\mu]} 
\lrprt_\alpha + c^{\alpha [\nu}\gamma_\alpha
{\stackrel{\leftrightarrow}{\partial^{\mu]}}}) \chi \\
\nonumber
& & -\frac i 2 \overline \chi (d^{[\nu \alpha}\gamma_5 \gamma^{\mu]} 
\lrprt_\alpha + d^{\alpha [\nu}\gamma_5 \gamma_\alpha
{\stackrel{\leftrightarrow}{\partial^{\mu]}}})
\chi + \frac i 2 e^{[\mu} \overline \chi
{\stackrel{\leftrightarrow}{\partial^{\nu]}}} \chi \\ 
& & - \frac 1 2f^{[\mu} \overline \chi \gamma_5
{\stackrel{\leftrightarrow}{\partial^{\nu]}}} \chi - \frac i 4
\overline \chi (2g^{[\nu \alpha \beta}\si^{\mu]}_{~\alpha}
\lrprt_\beta +g^{\al\be[\nu}\si_{\al
\be}{\stackrel{\leftrightarrow}{\partial^{\mu]}}}) \chi -
\overline \chi H^{[\mu \al}\si^{\nu]}_{~\al} \chi
\quad .
\eea
All of these equations can be verified by direct calculation.

It is expected that terms which can be eliminated using a field
redefinition should also be removable from $X^\mn$ with the
appropriate redefinition of the associated currents.
To explicitly construct the conserved currents, the field
redefinition is applied to the conventional currents written in terms
of $\psi$, satisfying the conventional Dirac equation.  
These conserved currents are re-expressed in terms of $\chi$ giving
the proper conserved currents for the new lagrangian. 

Alternatively, the conserved currents of the modified lagrangian may
be computed from N\"oether's theorem using a modified action of the
Lorentz group on the spinor fields.
Using a general finite field redefinition $\psi = e^{v \cdot \Gamma}
\chi$ , it is found that the action of U(1) and translations is
the same on $\psi$ and $\chi$ meaning that the 4-current and
energy-momentum tensors are computed as in
Eqs.(\ref{4cur}, \ref{emom}).
However, the action of the Lorentz transformations are in fact
modified due to the field redefinition.  
Under an infinitesimal Lorentz transformation of the
coordinates, the $\psi$ spinor components mix according to
$\delta\psi = S(\Lambda) \psi$, while the corresponding change in
$\chi$ is calculated to be $\delta \chi = e^{-v\cdot \Gamma}
S(\Lambda)e^{v \cdot \Gamma} \chi = \tilde S(\Lambda) \chi$.  This
means that the $\chi$ components transform according to a similar
representation of the complex Lorentz group.   When the associated
current is computed using this modified action of SL(2,C) on $\chi$,
it is indeed conserved.  

As an example, consider the field redefinition used in
Eq.(\ref{aeredef}) involving ${\rm Re} v_\mu$. If the conventional
$S(\Lambda)$ is used to find the current associated with the Lorentz
transformations, the result is
$j_\chi^{\alpha \mn}$ given by Eq.\rf{lorcur} which is not conserved.
However, if $\tilde S(\Lambda)$ is used to
transform the field $\chi$, then the associated current is 
\beq
\tilde j_\chi^{\alpha \mn} = x^{[\mu} \Theta^{\alpha \nu]} 
+{1 \over 4} \overline \chi \{ \Gamma^\alpha , \sigma^\mn\}\chi
+{1 \over 2} \overline \chi e^{[\mu}\sigma^{\nu] \alpha}\chi
\quad ,
\eeq
which is in fact conserved.
This shows that one must be careful to map the correct conserved
generators into the proper associated currents written in terms of
$\chi$.  Similar maps between generators can be performed using the
derivative transformations, however, in this case the $U(1)$ and
translation currents may also be modified.

As a practical approach, the redundant parameters can first
be removed from the lagrangian and then the conventional $S(\Lambda)$
can be used to construct the currents in the redefined lagrangian. 
We perform this removal of the redundant parameters in the next
section.

\vglue 0.6cm
{\bf \noindent V. ELIMINATION OF REDUNDANT PARAMETERS}
\vglue 0.4cm

In this section, we start with a general lagrangian and apply field
redefinitions to remove redundant parameters.  
The starting point is the lagrangian given in Eq.(\ref{lorviol}).  The
parameters $m_5$, $a_\mu$, $e_\mu$, $f_\mu$, and $c_{[\mn]}$ can all be
immediately removed using a combination of the
transformations described in section III.  
The parameter $g_{\lambda \mu \nu}$ can be replaced by
the traceless $\tilde g_{\lambda\mu\nu}$ satisfying $g^{\lambda
\nu}\tilde g_{\lambda\mu\nu} = g^{\mn}\tilde g_{\lambda \mn} = 0$. 
The other transformations listed in table 1 involve linear
combinations of parameters.  

To handle these terms, the lagrangian must be re-expressed in terms of
the new, physically relevant linear combinations of parameters.  For
example, the combination of antisymmetric $d$ and $H$ terms of
Eq.\rf{genlag} are re-expressed as
\bea
\frac i 4 \overline \psi d_{[\mn]} \gamma_5\gamma^\mu
\lrprtnu
\psi  - \frac 1 2 \overline \psi H_{\mn} \si^\mn \psi & = &
v^+_{\al\be} \left[\frac i 2\overline \psi \ep_\mn^{~~\al\be}
\gamma_5 \gamma^\mu \lrprtnu \psi - m \overline \psi \si^{\al
\be}\psi\right] \nonumber \\ 
& &+ v^-_{\alpha \beta}\left[\frac i 2 \overline \psi
\ep_\mn^{~~\al\be}
\gamma_5 \gamma^\mu \lrprtnu \psi + m \overline \psi \si^{\al
\be}\psi\right] \quad ,
\nonumber \\
\eea
where $v^\pm_{\alpha \be} = {1 \over 4}(\tilde d_{\al \be} \pm {1 \over
m} H_{\al \be})$ , and $\ep_\mn^{~~\al \be}\tilde d _{\al \be} =
d_{[\mn]}$.  The combination $v^-_{\al\be}$ can be removed using
a field redefinition, leaving only $v^+_{\al\be}$ terms in the
lagrangian.
The $g$ and $b$ terms can be similarly combined using the definition
$v^\pm_\al = \half (g_\al \pm \frac 1 m b_\al)$

After all of these redefinitions have been performed, the form for the
lagrangian becomes 
\bea
\label{redlag}
\cl = & &\frac i 2 \overline \psi (\gamma_\nu +\half c_{(\mn)}
\gamma^\mu + \half d_{(\mn)}\gamma_5 \gamma^\mu+ \half \hat g_{\la\mn}
\si^{\la\mu}) \lrprtnu \psi- m \overline \psi \psi \nonumber \\
& & +\frac 1 4 (\tilde d_{\al \be} + \frac 1 m  H_{\al \be})
\left[ \frac i 2 \overline \psi \ep_\mn^{~~\al\be} \gamma_5
\gamma^\mu \lrprtnu \psi - m \overline \psi \si^{\al \be} \psi\right]
\nonumber \\
& & - \frac 1 2 (g_\al - \frac 1 m b_\al)\left[ \frac i 2 \overline
\psi
\half \ep_{\la\mu}^{~~\nu \al}  \si^{\la \mu}\lrprt_\nu \psi - m
\overline
\psi \gamma_5 \gamma^\alpha\psi\right]
\quad ,
\eea
where $\hat g_{\la \mu \nu}$ is a traceless coupling with a vanishing
totally antisymmetric piece.  The totally antisymmetric component of
$\tilde g$ is absorbed into $g_\al = - \ep_\al^{~\la \mu \nu}\tilde
g_{\la\mu\nu}$.
This lagrangian can be written in the form 
\beq
\cl = \frac i 2  \overline{\ps} \tilde \Gamma_\nu \lrprtnu \ps 
- \overline{\ps} \tilde M \ps
\quad ,
\label{finallag}
\eeq
where 
\beq
\tilde \Ga_\nu = \ga_\nu +\half c_{(\mu \nu)} \ga^\mu 
+(\half d_{(\mn)} +\ep_{\mu\nu}^{~~\al\be}
v^+_{\al\be})\ga_5 \ga^\mu + \half (\hat g_{\la \mu \nu} -
\ep_{\la\mu\nu}^{~~~\al}v_\al^-)
\si^{\la \mu}
\quad ,
\eeq
and
\beq
\tilde M = m(1 + v_{\al\be}^+\si^{\al\be}
-v_\al^-\ga_5 \ga^\al)
\quad ,
\eeq
where all distinct parameters are now physically relevant.

This lagrangian is therefore the one for which $\psi$ can be assumed to
transform under the standard representation of SL(2,C), yielding
maximally conserved currents\footnote{Note that field redefinitions involving the 
physically relevant coupling constants may still
be performed yielding different definitions of the Lorentz currents.  However, the 
physically irrelevant parameters have been removed.  For example, a further field
redefinition is convenient in removing extra time derivatives for calculational
purposes as is described preceding Eq.\rf{trem}.}.  The relevant
currents are given as in the previous section with appropriately
mapped constants found by comparison of Eq.\rf{genlag} and
Eq.\rf{finallag}.  The remaining terms cannot be removed by a
redefinition of spinor coordinates. Other terms may be eliminated in
the free theory using  transformations involving derivatives of the
fields as is discussed in section III, but interactions between
the fermion fields and other fields often break the natural
correspondence between the redefined lagrangian and the original
theory.  This means that if these terms are removed from the free
fermion sector, they will appear as modified interaction terms and
will not be removed from the theory, only shifted to another sector.

Physical quantities
should therefore depend only on the combinations occurring in the
lagrangian of Eq.\rf{finallag}.  Therefore, it is not possible to
obtain experimental bounds on all Lorentz-violating parameters of
Eq.\rf{genlag} independently, only on the combinations present in
Eq.\rf{finallag}.  For example, it is only the linear combination
$v^+_{\al
\be}$ of the antisymmetric part of $d$ and $H$ that is observable. 
This implies that only one parameter should in fact be used to
describe this quantity.

Comparison with previous calculations within the context of this model
verify that this is indeed the case.  
For example, in applications to electrons and positrons in Penning
traps \cite{bkr,gg,hd,rm}, the relevant experimental bound is obtained
from the observable cyclotron and anomaly frequencies ($\tilde
g_{\la\mu\nu} = 0$ in this calculation)
\bea
& &\omega_c^{e^\pm} \approx (1 - c_{00}^e - c_{11}^e -
c_{22}^e)\omega_c
\quad , \nonumber \\
& & \omega_c^{e^\pm} \approx \omega_a \mp 2 b_{3}^e +2 m d_{30}^e
+2H_{12}^e
\quad ,
\eea
which can place bounds only on the combinations of parameters 
found in Eq.\rf{redlag}, not on the parameters that can be removed by
the field redefinitions.  
As another example, a calculation of the cross-section for $e^+ e^- \rightarrow 2 \ga$
within the QED extension \cite{scatt} only depends on the symmetric
components of $c$. Similar results have been obtained in other QED
systems
\cite{robertrev}.

For many practical calculations it is convenient to perform another
field redefinition to normalize $\tilde \Gamma^0\rightarrow \gamma^0$.
This has the effect of removing extra time derivative couplings
insuring that the  resulting
Schr\"odinger Equation has a conventional time evolution
\cite{bkr, kla}.
Starting with a general lagrangian of the form \rf{lorviol}, the
appropriate field redefinition that removes time derivative couplings
to lowest order is
$\psi = A\chi$ with
\bea
A & = & 1 - \half\ga^0(\Ga_0 - \ga_0) \nonumber \\
& = & 1 - \half\ga^0(c_{\mu 0} \ga^\mu +d_{\mu 0}\ga_5 \ga^\mu+
e_0 + i f_0 \ga_5 + \half g_{\la\mu 0}\si^{\la\mu}) \quad .
\label{trem}
\eea
Note that these are in 1-1 correspondence with the non-unitary
transformations of the form $v \cdot \Ga$ examined in section III.
The unitary transformations of this type can be obtained by letting
the coupling constants in Eq.\rf{trem} be pure imaginary rather than
real.  In this case, the time derivative structure is unaffected by
the field redefinition.  This provides an alternative perspective on
the spinor component field redefinitions.  The unitary transformations
preserve the time derivative terms while the non-unitary
transformations modify the time derivative structure.  This can be
seen directly from Eq.\rf{vdotgamlag} by noting that the requirement
that no time derivatives are introduced is equivalent to the condition
$(v \cdot \Ga)^\dagger = -v \cdot \Ga$, hence making the field
redefinition unitary to lowest order.  
The unitary transformations are used in the standard Dirac theory to
transition between various representations of the gamma matrices, an
alternative perspective to the explicit correction terms 
used in this paper. 
In other words, any apparent Lorentz-violating terms generated by a
unitary transformation may be absorbed into a modified gamma matrix
representation.

As a practical method for calculation, the procedure is therefore to
first remove the redundant parameters to obtain Eq.\rf{finallag}, then
perform the field redefinition of Eq.\rf{trem} (using $\tilde \Ga^0$ in
place of $\Ga^0$).
The resulting lagrangian will therefore yeild a conventional
Schr\"odinger Equation time evolution and will not contain any
redundant parameters.

\vglue 0.6cm
{\bf \noindent VI. SUMMARY}
\vglue 0.4cm

In this paper, an analysis of field redefinitions in
the context of the Lorentz-violating QED extension was presented.  It
was shown that a variety of parameters that apparently violate Lorentz
invariance can be eliminated from the lagrangian using suitable
fermionic field redefinitions.   

The approach taken to find these parameters was to begin with the
conventional Dirac lagrangian, introduce an arbitrary spinor
redefinition, and examine the resulting transformed lagrangian.  Any
parameters generated using this procedure can be removed by the
corresponding inverse transformation.  The action of
SL(2,C) on the spinor fields is deduced using the conventional action
on the $\psi$ spinors and performing the field redefinition to
determine the corresponding action on
$\chi$.  The resulting transformation matrices of the new spinor are
related by a similarity transformation to the original matrices.  This
modified action must be taken into account when defining the conserved
generators in the modified lagrangian.  

This implies a procedure for elimination of redundant parameters from
the original lagrangian.  By first identifying all possible terms that
can be generated from the conventional lagrangian by a field
redefinition, these terms can be omitted from the physical
lagrangian. This procedure has been carried out in section V of
the paper.
A further transformation may be implemented to normalize the time
derivative structure of the theory to obtain conventional
Schr\"odinger time evolution and physical particle states.

The possible nonderivative field redefinitions that can be applied to
the QED extension fall into two general categories, unitary and
nonunitary.  The nonunitary transformations
modify the time derivative structure of the lagrangian and can be used
to construct a hermitian hamiltonian and a
Schr\"odinger equation with conventional time evolution.  The time
derivative couplings in a general bilinear lagrangian of the form in
Eq.\rf{lorviol} can always be removed using a suitable nonunitary
transformation.  Moreover, there is a one-to-one correspondence
between these nonunitary transformations and the field redefinitions
used to eliminate the extra time derivative couplings.  

Redefinitions involving differentiation may be useful in the free fermion
theory, but often lead to nonlocal interactions or skewed coordinate
systems when interactions are present.  These problems make it
difficult to perform a generic analysis of all possible applications
of these transformations.  Derivative redefinitions may be applied on a
case by case basis where they might be useful in simplification of
calculations.

Stability and causality issues \cite{kle} cannot be effectively
addressed using the above arguments since the redefinition was only
carried out to lowest order in Lorentz-violating
parameters.  Causality and
stability problems appear either when the coupling constants are
large, or when the momentum is significantly large to invalidate the
linear approximations involved.  However, the finite field redefinition
considered in Eq.\rf{finitefredef} leads to a finite set of parameters
that maintain the conventional dispersion relation.  The resulting
theory must therefore be stable and microcausal.  A class of apparently
Lorentz-violating lagrangians that are microcausal and stable may 
therefore be generated by applying finite
versions of the field redefinitions discussed in this paper.  A
complete nonperturbative analysis would be of interest, but is beyond
the scope of this work.

Application of a similar analysis to the entire standard model
extension would be of interest. 
For example, cross-generational mixings would allow for a richer
structure of possible field redefinitions than in QED.

\vglue 0.6cm
{\bf \noindent ACKNOWLEDGMENTS}
This work was supported in part by a University of South
Florida Division of Sponsored Research grant.  We thank Alan
Kosteleck\'y and Robert Bluhm for discussions.

\vglue 0.6cm
{\bf\noindent REFERENCES}
\vglue 0.4cm

\end{document}